# Electrical analysis of hysteresis in solution processed silicon nanowire field effect transistors

K. Prabha Rajeev, C. Opoku, V. Stolojan, M. Constantinou and M. Shkunov*

Electronic Engineering, Advanced Technology Institute, University of Surrey, Guildford GU2 7XH, UK

**Abstract**   Silicon nanowires (Si NW) are ideal candidates for solution processable field effect transistors (FETs). The interface between the nanowire channel and the gate dielectric plays a crucial role in the FET performance, and it can be responsible for unwanted effects such as hysteresis of the I-V characteristics due to threshold voltage shift when the gate voltage is applied. Using gate-voltage bias stress measurements we show that a large hysteresis of up to 40V in Si NW FETs with $SiO_2$ dielectric is mainly due to the holes traps at the nanowire/$SiO_2$ interface.  An approach for reducing this hysteresis to just 2.5V using solution processable fluoropolymer dielectric Cytop in the top-gate configuration is demonstrated. Experimental results suggest that the density of surface traps in Si nanowire transistors is dictated mainly by the nature of the dielectric layer. The influence of the gate dielectric was studied by assessing the field effect transport behaviour of a representative double gate FETs based on $SiO_2$ bottom dielectric and top a Cytop dielectric layer. Such devices were characterised, revealing an order of magnitude higher hole traps density at the nanowire/$SiO_2$ interface ($1\times10^{13} cm^{-2}$) compared to that of nanowire/fluoropolymer interface ($7.5\times10^{11} cm^{-2}$).

* email:  m.shkunov@surrey.ac.uk

1. **Introduction**

Solution-processed printable electronics using semiconducting inks have opened up the possibility of various low-cost electronic devices targeting large area applications such as RFID tags[1], sensor[2], displays[3] where semiconductor is deposited via various printing methods compatible with a wide range of substrates including plastics. Currently, the materials of choice for large area electronics are amorphous silicon (a-Si), polycrystalline silicon (poly-Si) and organic semiconductors. However, the high temperatures required for processing of amorphous silicon or polysilicon transistors make it very challenging to use them on low-cost plastic substrates[4]. Conversely, organic semiconductors can be deposited at low temperatures, but their charge carrier mobility in printable FETs is limited to few $cm^2/Vs$[5]. Nanomaterials such as inorganic semiconducting nanowires offer very high mobilities[6], and their low temperature processing[7] makes them ideal candidates for printable electronics. Growth methods such as chemical vapour deposition growth[8], vapour-liquid-solid growth[9] can produce high quality nanowires, with lengths compatible with typical transistor channel dimensions.  However, such synthesis methods are limited to small substrate size thus making them impractical for industrial scale processes.  Korgel's group[10] has demonstrated that the supercritical fluid liquid solid (SFLS) method for growing Si nanowires can be industrially scalable, with a throughput of few kilograms per day is achievable. The ability to disperse these nanowires in different solvents to make ink formulation shows their potential for printing deposition technologies. The separation of the synthesis process from the devices assembly enables the fabrication of electronics at low processing temperatures.

In order to fully realise the potential of solution processable Si NWs in the area of high performance transistors, some of the key issues need to be addressed such as: i) alignment of Si nanowires to form 'monolayer' of ordered arrays to bridge the source-drain electrodes, and ii) elimination of surface states to reduce I-V characteristics hysteresis and increase the device mobility.

Various alignment deposition techniques, such as directed flow assembly[7], DNA hybridisation assembly[11], electrostatic force assembly[12], Langmuir- Blodgett technique[13, 14], dielectrophoresis alignment[15, 16] have been demonstrated, however, the majority of these techniques cannot be applied to large area deposition or they involve laborious fabrication steps to do so[7]. A high quality alignment, where nanowires do not cross each other is essential to avoid gate screening effect for high performance FETs.  Thus, a low-cost and large substrate area compatible nanowires alignment techniques are still required.

Nanowires possess high surface area, and charge transport is strongly affected by the quality of semiconductor surface. In particular, the device mobility and hysteresis in Si NW FETs are influenced by the nanowire/gate-dielectric interface, due to the trapping of charge carriers. Commonly used gate dielectric for Si NW FET fabrication is thermally grown silica ($SiO_2$). $SiO_2$ dielectric is known to contain surface bound hydroxyl ($OH^-$) groups[17, 18]. These groups induce interface states that can trap accumulated charges during FET characteristics. The trapping is typically manifested as hysteresis behavior during transistor I-V scans. Hysteresis is the shift in threshold voltage ($\Delta V_{th}$) during the forward (+V to –V) and the reverse gate voltage sweep (-V to +V) at constant drain voltage bias. The mobility in a-Si FETs using *high-k* gate dielectrics is compromised due to scattering mechanism from fixed charges[19], and an increase in hysteresis is reported for Si NW FETs due to the surface adsorbed $OH^-$ groups[18]. Elimination of hysteresis in a device will improve the overall FET performance including the device field effect mobility, due to less scattering of charge carriers. Passivation of devices with poly (methyl methacrylate) on PbSe nanowires or self-assembled monolayers on Si NWs has shown to remove the surface bound hydroxyl group, thereby reducing the hysteresis. It has also been reported that $SiO_2$ thermally grown at high temperatures (~1100ºC) on Si NW reduces the hysteresis in the corresponding nanowire FETs, but such high temperature treatment is not compatible with flexible substrates [20, 21]. Gate dielectrics which have low affinity towards $OH^-$ groups such as silicon nitride (*a*-$SiN_x$) have shown increased mobility in *a*-Si thin film transistors due to less interface charging[22], but the high temperature required for the deposition of the *a*-$SiN_x$ (~ 250ºC - 450ºC) makes them incompatible with plastic substrates. However, low k insulators such as organic dielectrics, have surfaces with low energetic disorder which were investigated in organic field effect transistors[23]. Reduced hysteresis and enhanced mobility can be achieved by using hydrophobic polymers as gate dielectrics instead of $SiO_2$, with low amount of $OH^-$ trap sites at the nanowire/dielectric interface[24].

In this work, we have deposited SFLS grown Si NWs via a roll-cast technique to give ordered arrays of nanowires and to remove nanowire aggregates and impurities clusters from the nanowire layers. Gold electrodes were used as source/drain contacts in the Si NW FETs to provide near-ohmic contacts. The nature of traps causing hysteresis was studied using sweep rate study and bias stress measurement of transistors with $SiO_2$ as dielectric, in bottom-gated geometry. The effect on hysteresis using a fluoropolymer (Cytop with $\varepsilon_r$ ~ 2.1) as the gate dielectric was studied in top gate device geometry. Finally a comparison of $SiO_2$ and fluoropolymer as gate dielectrics in a dual-gate configuration of the same FET device was conducted showing significant reduction of surface trap density at nanowire/polymer interface compared to nanowire/silicon oxide interface.

## 2. Experimental details

Silicon nanowires used in this work were synthesised by the supercritical-fluid liquid solid method, as described elsewhere[10]. The formulations of Si NWs dispersed in anisole were used to prepare samples for SEM and TEM analysis by drop casting a small amount of dispersion on Si/$SiO_2$ substrates and on a holey-carbon grid (Agar Scientific), respectively. SEM images of as-grown Si nanowires were used to evaluate the NW lengths to be from 2µm to 100µm. Most of nanowires were within 5-30µm length range. Figure 1(a) shows TEM image of a network of Si nanowires with a noticeable amount of impurities, nanowire aggregates and defective kinked nanowires. High resolution TEM in Figure 1(b) show a typical straight single nanowire with a diameter of ~ 35nm surrounded by a 4nm thick amorphous polyphenylsilane shell. The nanowire growth direction was identified from computed Fast Fourier Transform (FFT) pattern obtained from the lattice fringes (Fig. 1(b)) to be along [110], which is consistent with previous reports[10].

Nanowires were deposited on the substrates using a roll-cast coating method. The nanowire formulation was drop-casted on the part of the inclined substrate, and a glass roller was placed at the top of the substrate and then allowed to roll down under the influence of the gravitational force. The roller was making contact with the substrate, enabling to spread out the nanowire formulation and to remove protruding clumps, stacked nanowires and large impurity particles. Additionally, the shear force induced by the roller movement was acting on nanowires to produces preferential alignment along the rolling direction (supporting information, Figure S1). The nanowires density on the substrate can be increased by repeating the process, until the required density is achieved. The nanowire alignment can be improved by using a compressed air gun to dry the solvent immediately after the roll-casting process. Following the nanowire deposition the substrates were dried and prepared for photolithography.

The source-drain transistor electrodes were patterned using photolithography (lift-off) on top of the aligned nanowires by sputtering Cr/Au (3nm/50nm) contacts. Bottom-gate FETs were fabricated on Si/$SiO_2$ substrates, whereas top-gated FETs were prepared on low sodium glass substrates. Following the contacts deposition the substrates were annealed at 200°C in $N_2$ filled glove box, to improve charge injection/extraction at the nanowire-contact interface. For top-gate transistor devices fluoropolymer Cytop was spin coated on top of deposited nanowires with source-drained patterned electrodes; and then baked (100°C, 10min) in air. A 50nm thick Au gate electrode was evaporated through a shadow mask on top of the channel area covered with gate dielectric. The double gate devices were fabricated by preparing a bottom gate FET on $Si/SiO_2$ substrate and then completing the structure with a top gate insulator (Cytop) and a gate electrode.

Transistor characterisation and the gate-voltage bias stress measurements were performed with Agilent 4155C semiconductor analyzer in $N_2$ filled glove box to minimize the environmental effect which can cause traps due to surface bound $OH^-$ groups and adsorption of polar water molecules onto the nanowire or nanowire/dielectric interfaces.

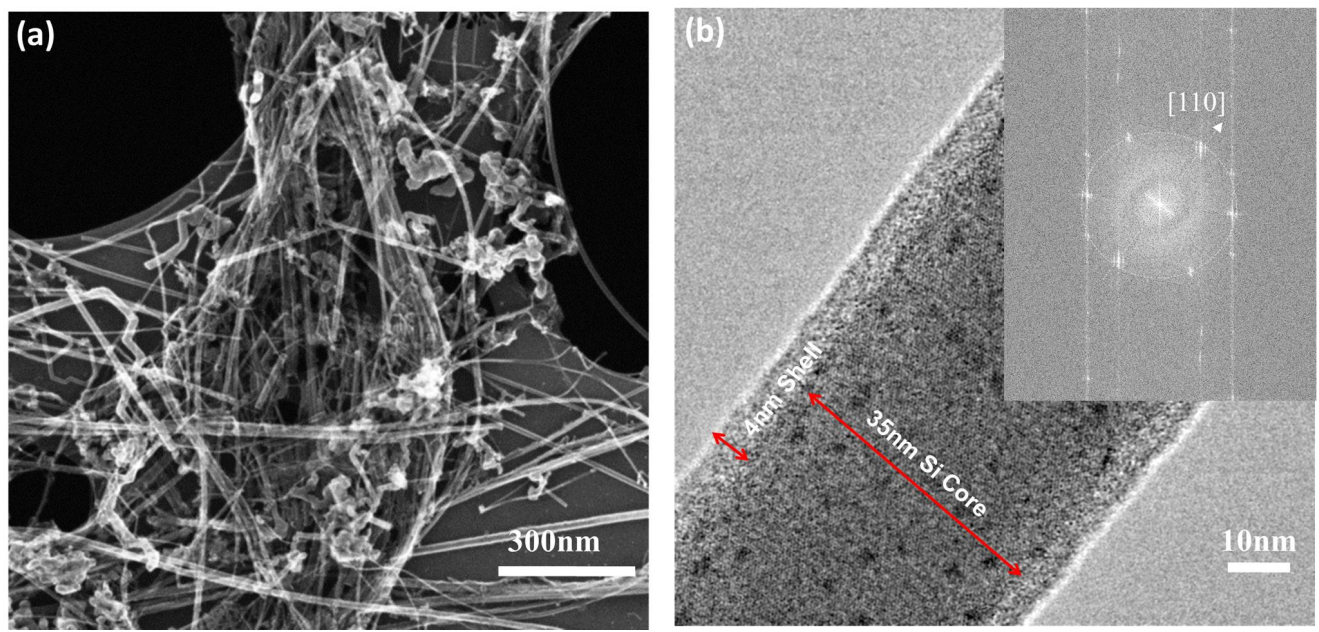

**Figure 1.** Characterisation of SFLS-grown silicon nanowires. (a) TEM secondary electron image of a network of nanowires showing impurity particles and clumps of nanowires. TEM carbon grid is visible on the background, (b) High resolution TEM (HRTEM) image of a single nanowire with 35nm core diameter and 4nm thick amorphous shell. The inset illustrates FFT of the nanowire showing the growth direction [110].

## 3. Results and discussion

### 3.1 Bottom gate vs top gate Si - NW FETs

Figure 2(a) shows the transfer characteristics of the Si-NW FETs (23 nanowires in the channel) with $SiO_2$ as gate dielectric, with channel length of 5μm. The drain current ($I_D$) was obtained by scanning the gate voltage from +20V to -60V in the forward voltage sweep and from -60V to +20V in the reverse voltage sweep, at a -5V and -10V drain bias voltages ($V_D$). The gate voltage sweep rate was kept constant at 1 V/s throughout the measurement. A lower gate voltage sweep rate is chosen to study the extent of trapping and trap depletion at the Si NW/$SiO_2$ interface. In our studies, we have observed that the hysteresis increases with decreasing gate sweep rate, indicating the presence of slow filling traps. From the transfer characteristics in Figure 2 (a), a p-type behavior is observed showing charge accumulation at negative bias gate voltages. The device shows an excellent gate modulation for a common back-gated transistor with an $I_{ON}/I_{OFF}$ ratio of $10^6$, and a turn ON voltage at -2V. An increase in $I_{ON}$ with $V_D$ indicates good gate to channel coupling. The hysteresis ($\Delta V_{th}$) was extracted by taking the difference of threshold voltages obtained from forward gate sweep current ($I_D$) and reverse sweep (Supporting Information (SI) Fig.S2). A large hysteresis of ~ 30V was observed at -5V drain voltage whereas the hysteresis decreased to ~ 24V at -10V drain bias (Fig.2(a)). The decrease in hysteresis at high $V_D$ can be attributed to the faster de-trappings of charge carriers at higher lateral electric field. A subthreshold swing (SS

= $\Delta V_G/\Delta \log I_D$) of 1.8V/decade and transconductance (Gm = $\partial I_D/\partial V_G$) of 0.6μS were estimated from the transfer characteristics.

The transfer characteristics of the top-gated Si-NW FETs with fluoropolymer dielectric are shown in Figure 2 (b). The device has a channel length of 2.5μm with 4 nanowires bridging the source-drain electrodes. The hysteresis is reduced significantly more than an order of magnitude, when compared to bottom gate Si NW device with SiO$_2$ dielectric. The measured hysteresis is ~ 2.3V at -5V drain bias, which indicates less trapping of charge carriers at the Si NW/polymer dielectric interface. There is no significant change in hysteresis when measured at a higher drain bias voltage (V$_D$ -10V). High I$_{ON}$/I$_{OFF}$ ratio of > 10$^7$ is obtained with transconductance of 0.6μS. The subthreshold swing (SS) was estimated to be ~ 3.3V/decade. The higher SS for top gate Si-NW FETs is due to the larger thickness of Cytop (1μm) compared to 230nm thick SiO$_2$ for bottom-gated devices, and also lower dielectric constant.

To extract charge carrier mobility the parasitic capacitance and fringing of the gate field due to the cylindrical nature of the nanowires should be taken into consideration. The nanowire FET mobility was estimated in the linear regime using the following equations [25-27]:

$$\mu = Gm \times \frac{L^2}{V_D C_n} \quad (1)$$

$$C_n = N \times \frac{2\pi\varepsilon_o\varepsilon_{ins}L}{\cosh^{-1}(\frac{r+d}{r})} \quad (2)$$

Where $C_n$ is the capacitance based on cylinder on a plate model for *N* number of nanowires in the FET channel with radius *r* and gate dielectric thickness *d*. *L* is the channel length, *ε$_o$* is the absolute permittivity, *ε$_{ins}$* is the gate insulator dielectric constant and V$_D$ is the drain bias voltage. A capacitance of ~ 7.2× 10$^{-15}$ F was calculated for Si-NW FETs on SiO$_2$ using equation (2), whereas capacitance of Si-NW/Cytop structure was ~ 2.5× 10$^{-16}$ F. Using equation (1), we have calculated that the Si-NW FET mobility increased from 8 cm$^2$/Vs for the bottom gate device to 14 cm$^2$/Vs for top-gate device with Cytop dielectric. The increase in mobility for Si-NW/Cytop FETs is due to the less traps present at the interface, reducing the scattering of charge carriers. The change in hysteresis was compared for SiO$_2$ and organic polymer dielectric measured for several devices and they all showed similar pattern. (See Supporting Information, Figure S2). The hysteresis values for SiO$_2$ dielectric FETs were up to 40V, and typical ΔV$_{th}$ values were in the range of 15V to 40V (SI Fig.S3). Devices with Cytop dielectric had significantly lower hysteresis, below 10V, with 50% of devices showing ΔV$_{th}$ of less than 3V.

The nature of traps in the SFLS-grown Si-NW FETs was analysed using gate-voltage bias stress measurement, as discussed in the next section.

## 3.2 Gate voltage bias stress measurement

Gate voltage bias stress experiments was carried out to determine if the observed hysteresis in Si-NW FETs is due to majority carrier (holes) trapping or minority carrier (electrons) trapping at the Si-NW/dielectric interface. The transfer characteristics of Si-NW/SiO$_2$ and Si-NW/Cytop FETs were initially measured before the gate voltage bias stress measurements. Then the gate electrode was biased at a constant -40V voltage for 30mins, and transfer characteristics were immediately measured, at a drain bias of -10V. The stress cycles were repeated for a total of 12hrs, with 30min stress time periods followed by transfer characteristics measurements. The stress measurements with +40V gate bias voltage was performed after the negative voltage bias stress, with the same stress measurement cycles for 12hrs. Figure 2(c) shows the extracted hysteresis values from the transfer characteristics of Si-NW/SiO$_2$ FET, for the gate bias stress measurements. It is observed that the hysteresis decreased to ~ 19.3V after the -40V gate bias compared to an initial value of ~ 29.1V, whereas the hysteresis increased from 24.4V to ~ 31.4V after +40V gate voltage stress. The threshold voltage also shifts to a further negative values (~ -12V) after the -40V stress voltage compared to the initial threshold voltage of ~ -16V, however, the threshold voltage after +40V bias stress has decreased to ~ -2V (see SI, Figure S4). A lower hysteresis is observed after the negative gate-voltage stress due to the trapping of holes, thereby filling the long-lived traps with charge carriers. This experiment indicates that most of the traps will be occupied with holes during the 30min hold time of -40V gate-voltage stress, resulting in a higher reverse I$_D$ and lower hysteresis. Whereas during positive gate-voltage bias stress, holes get de-trapped. This effect results in density increase of unoccupied traps present during the following I-V measurements, thereby increasing the hysteresis after +40V gate-voltage bias stress.

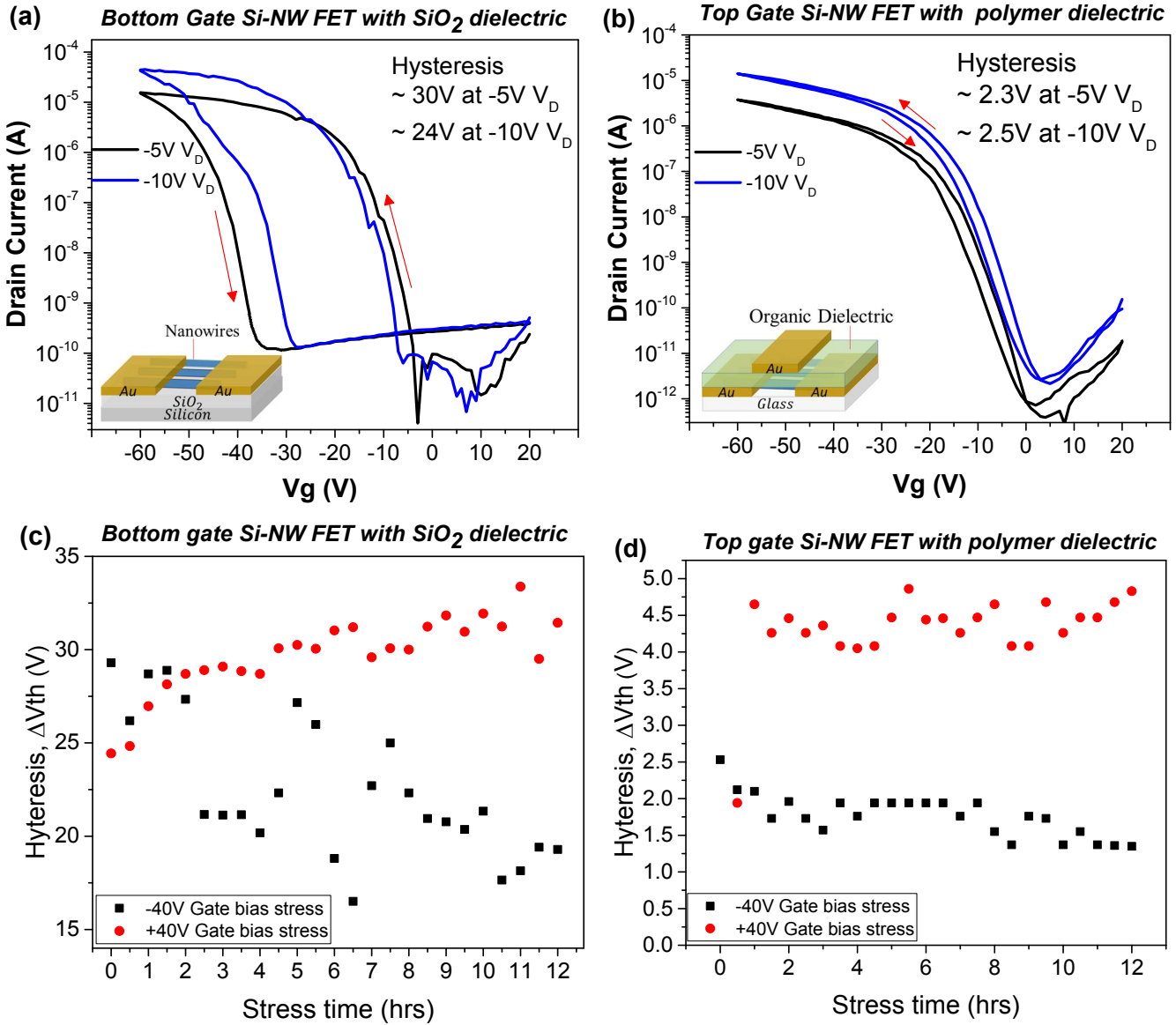

**Figure 2.** (a,b) transfer characteristics of Si NW FETs: (a) bottom gate FET with SiO$_2$ dielectric with 23 nanowires in the channel, showing a hysteresis value of ~30V at -5V drain voltage, with the inset showing the device architecture, (b) top gate FET with fluoropolymer dielectric with 4 nanowires in the channel, showing a hysteresis value of ~2.3V at -5V drain bias, with the inset showing the device architecture, (c) and (d) extracted hysteresis from the bias stress measurement for bottom gate SiO$_2$ dielectric FET and top gate devices with polymer dielectric, at -40V gate bias stress for 12hrs and +40V gate bias stress for 12hrs.

If we consider the hysteresis change and the shift of the threshold voltage after the bias stress in Si NW/SiO$_2$ FETs, these effects can be explained by the presence of both shallow and deep traps at the nanowire/SiO$_2$ interface. The shallow traps cause a change in hysteresis, whereas the deep traps, which take longer time to empty, causes the shift in the threshold voltage during the bias stress measurements[28]. Therefore, the shift in the threshold observed towards negative gate voltage during the -40V gate-voltage bias stress indicates the presence of deep hole traps at the Si NW/SiO$_2$ interface which screen the gate electric field. The positive shift in the threshold voltage during +40V gate bias indicates the de-trapping of holes, followed by possible electron trapping. However, the electron trapping is not significant as the threshold voltage shift is almost constant after the first stress period (30min). The absence of steps or 'knees' in the subthreshold FET characteristics during the negative stress conditions indicates that no additional traps are formed (SI, Figure S4) [22, 29]. Therefore the main contributors of the strong hysteresis in Si NW FET using SiO$_2$ dielectric are the majority (holes) carrier traps at the Si NW/SiO$_2$ interface.

Figure 2(d) shows the extracted hysteresis from the transfer characteristics obtained after positive and negative gat-voltage bias stress for the top gate Si-NW/Cytop FET. The hysteresis decreased to ~ 1.3V from an initial value of ~2.5V after -40V gate voltage bias stress, and an increase in hysteresis to ~ 4.8V was observed after +40V gate voltage bias stress measurement. The observed change in hysteresis shows similar trend to Si-NW/SiO$_2$ FETs, indicating the filling and emptying of hole traps. The change in hysteresis is not that significant as compared to devices with SiO$_2$ dielectric, which suggests that lower density of traps is present at the Si-NW/Cytop interface. A shift in the threshold voltage to ~ -22V was observed after -40V gate-voltage bias stress compared to an initial value of ~ -13V, whereas, a threshold voltage of ~ -5V was achieved after +40V gate-voltage bias stress (SI, Figure S5). The shift is constant after the first stress period for the positive bias indicating the absence of new electron traps introduced during the positive gate stress.

Overall, from the bias stress measurements conducted for Si nanowire FETs with SiO$_2$ and fluoropolymer dielectrics, we conclude that the majority of traps at the SFLS-grown Si-NW/dielectric interface are due to the shallow and deep hole traps, with significantly lower effect of trapping in the polymer top-gate dielectric devices, resulting in an order of magnitude reduction in hysteresis.

### 3.3 Double gate Si-NW FET

The direct one-to-one comparison of top gate and bottom gate devices presented in section 3.1 to study the NW/dielectric interface was not possible due to the differences in nanowire numbers present in the FET channels, and also due to possible variation in nanowires' properties including diameters, shell thicknesses, levels of impurities, and even Si core growth directions, resulting from non-uniform characteristics of as-synthesised SFLS nanowires[10, 30]. To be able to demonstrate the identical nanowire channel for bottom and top-gate transistor devices we have fabricated dual-gate FETs produced on the same substrate. The nanowires forming the FET channel were gated on the bottom by the SiO$_2$ dielectric, and on the top by the Cytop dielectric. These dual-gate structures allowed us to eliminate the variation of Si NW morphology, placement and the number of nanowires for the comparative studies of hysteresis effects in SiO$_2$ vs organic dielectric FETs. Figure 3 shows the transfer and output characteristics of a double-gate Si NW transistor with SiO$_2$ as bottom-gate dielectric and Cytop as top-gate dielectric, with 23 nanowires bridging the channel. Both the bottom gate and the top gate devices demonstrated excellent gate modulation, with an $I_{ON}/I_{OFF}$ ratio of ~$10^7$ for the bottom gate device and $3 \times 10^6$ for top gate Cytop device.

The hysteresis of the top gate organic dielectric device (~ 6V) was much lower compared to the hysteresis of the bottom gate SiO$_2$ device (~ 31V) (Fig.3a), due to lower trap density at the NW/organic dielectric interface, as we discuss below. These results are fully consistent with the measurements in section 3.1.

A higher $I_{ON}$ at $V_D$ ~ -5V was observed for the bottom gate SiO$_2$ device compared to the top gate organic dielectric device, due to the higher dielectric value for the low thickness (230nm) of the SiO$_2$ insulator ($\varepsilon_{ins}$=3.9) compared to the value of fluoropolymer dielectric ($\varepsilon_{ins}$=2.1), for 1μm thick film. The extracted Si NW FET mobility of the top gate Cytop device of 12 cm$^2$/Vs is significantly higher than mobility of the bottom gate SiO$_2$ device of 1.7 cm$^2$/Vs. The enhanced mobility for the top gate device can be attributed to the less scattering from trapped charge carriers in the Si NW/polymer dielectric interface.

The devices showed a near-Ohmic contact behavior, as observed in the output characteristics in Figure 3 (b,c). When both bottom and top gate were biased at the same time, there was an increase in $I_{ON}$ and the hysteresis value (~ 22V) was between the values obtained when biasing just the top and bottom gates individually (SI, Figure S6). The increase in $I_{ON}$ can be due to the formation of the double channel around the nanowire at both the interfaces, resulting in more charge formation and better performance. When both the gates are biased, the double gate device acts as a gate all around FET to give a higher mobility and an enhanced performance [11, 31]. Charge accumulation at the same voltage is higher at Si NW/SiO$_2$ interface when compared to Si NW/Cytop interface, due to the higher capacitance of SiO$_2$ compared to Cytop.

We have evaluated the occupied trap density at the nanowire-dielectric interface related to the hysteresis using the following equation[18]:

$$NqQ_t = \frac{C_n \times |\Delta Vth|}{2\pi r L_{NW}} \qquad (3)$$

Where, $q$ is the elementary charge and $Q_t$ is the occupied trap density. Other notations are defined in Eq. 1-2. Typical radius ($r$) of Si nanowires is ~ 20nm (± 5nm) and the nanowire length is estimated as the channel length (5μm) of the dual gate FET.

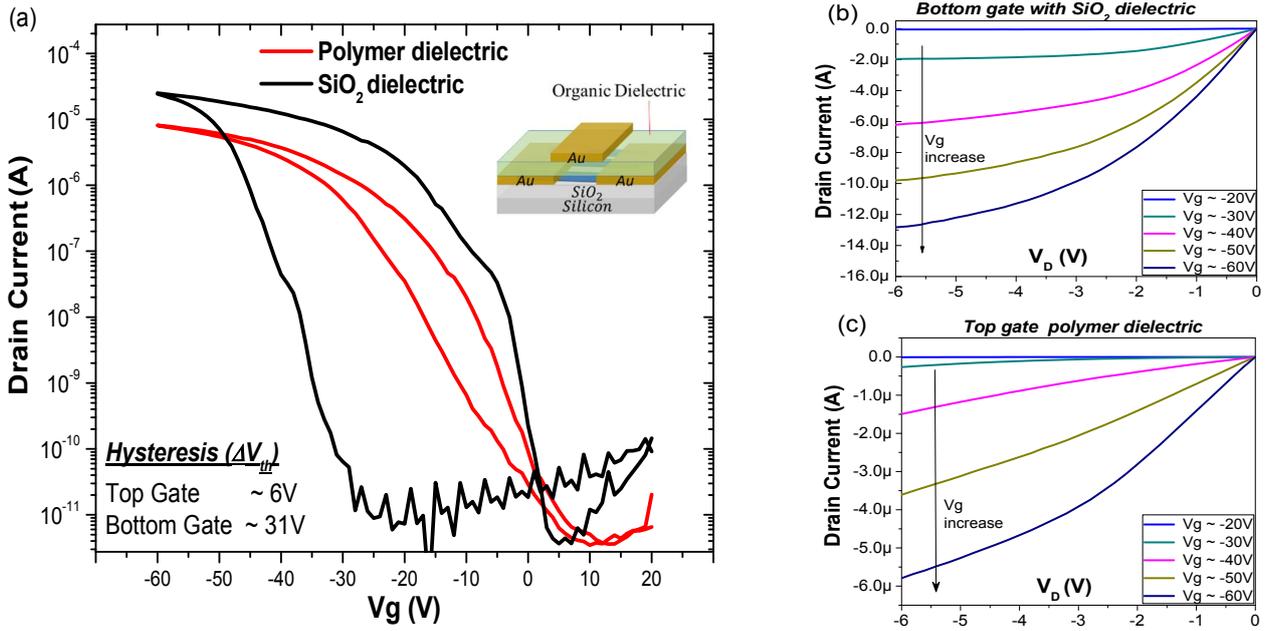

**Figure 3.** Transfer and output characteristics of the double gate Si NW FETs with SiO$_2$ bottom gate dielectric and fluoropolymer top gate dielectric: (a) transfer plots measurements of top gate and bottom gate devices (V$_D$ -5V); (b-c) output characteristics of bottom gate and top gate FETs obtained by scanning drain voltage from 0 to -6V and measuring the drain current at constant gate voltages [-20V, -30V, -40V, -50V, -60V].

For the bottom gate SiO$_2$ operation FET, $Q_t$ was evaluated to be ~ 1 × 10$^{13}$ cm$^{-2}$ which is more than an order of magnitude higher compared to the Si NW/Cytop top gate operation FET $Q_t$ ~ 7.5 × 10$^{11}$ cm$^{-2}$. This difference shows that the Si NW/organic dielectric forms a higher quality lower-trapping interface compared to the Si NW/SiO$_2$ interface. Even lower trap density of 2.8×10$^{11}$ cm$^{-2}$ was evaluated for the top-gate FET device prepared on glass substrate with 2.3V hysteresis, demonstrated in section 3.1 (Fig. 2(b)).

Lower I$_{OFF}$ of 3.6 × 10$^{-12}$A was observed for the forward scan of Si-NW/SiO$_2$ FET when compared to 1.5 × 10$^{-11}$A for reverse scan, demonstrating the effect of traps on the band bending at the NW interface. Higher band bending leads to higher Schottky barrier at the source/drain contacts, which will lower the I$_{OFF}$[18]. The subthreshold swing (SS) for the bottom gate FET was extracted to be 2V/decade, whereas the top gate device gave an SS of 4.3V/decade. The lower SS for bottom gate device is due to the higher dielectric constant of SiO$_2$ compared to the organic dielectric. It needs to be taken into account that the thickness of SiO$_2$ is ~ 4 times lower than the organic dielectric (1μm). The increase in reverse SS compared to forward SS for bottom gate FET (from 2.7V/decade to 5.1V/decade) indicates the downward band bending inside the Si NW to the field near the SiO$_2$ interface, which depends on the trap density[18]. Top gate devices showed a small shift in SS for forward and reverse scans (5.6V/decade to 4.9V/decade) with negligible shift in I$_{OFF}$ (2 × 10$^{-12}$A to 1.6 × 10$^{-12}$A) confirming the presence of less traps.

Finally, the standard (-40V) gate voltage bias stress was performed separately for the SiO$_2$ gate and the Cytop gate parts of the dual-gate FET. The hysteresis reduced from an initial value of ~ 45V to ~ 40V for SiO$_2$ dielectric after stress measurement, whereas Cytop dielectric gave ~ 8V hysteresis following the stress measurements, compared to an initial value of ~12V (SI, Figure S7). A shift in threshold voltage to negative values were also observed. The threshold voltage in SiO$_2$ dielectric FET shifted to ~ -27.5V compared to an initial value of ~ -21.4V, however, Cytop dielectric showed a significant shift in threshold voltage from an initial value of ~ -27V to ~-47V after the stress measurement. A reduced hysteresis and negative voltage shift in V$_{th}$ observed for both the SiO$_2$ and Cytop devices, indicates the presence of shallow and deep hole traps at the interface.

## 4. Conclusions

Solution-processed silicon nanowires were deposited and aligned with low impurity content (NW clumps) on top of substrates using a roll-cast technique. Nanowire field-effect transistor devices in bottom– and top-gate configurations with SiO$_2$ and Cytop polymer gate dielectrics were fabricated. Electrical analysis of the FETs demonstrated dramatic reduction of hysteresis from up to 40V in SiO$_2$ gated devices to only 2.3V in Cytop-gated transistor. Both individual bottom gate (SiO$_2$) and top-gate (Cytop) FETs, as well as dual-gate devices show the same trend of very low hysteresis for the hydrophobic polymer gate dielectric. The change in hysteresis behavior is linked with over 10 times lower occupied trap density at the nanowire-fluoropolymer dielectric interface of 2.3×10$^{11}$ cm$^{-2}$ (Cytop) vs 1 × 10$^{13}$ cm$^{-2}$ (SiO$_2$). Gate bias stress studies confirmed that the trapping of holes is taking place, as shown by the lower hysteresis (~ 19.3V) after 12hr negative (-40V) gate voltage stress, but an elevated hysteresis (~ 31.4V) after +40V bias, also indicating the presence of fast filling shallow traps at the nanowire-dielectric interface. The shift in threshold after the prolonged positive and negative gate bias voltages indicated the presence of deep, slow filling hole traps which screen the gate electric field. Finally, higher Si NW FET mobility of 12 cm$^2$/Vs observed for Cytop dielectric device compared to 1.7 cm$^2$/Vs for Si-NW/SiO$_2$ FET confirms the lower degree of charge carrier scattering at the nanowire/polymer interface, fully consistent with the lower trap density.

The demonstrated silicon nanowire transistors with solvent based processing of both semiconducting channel and dielectric layer open up possibilities for low-cost printing technologies for nanomaterials- based electronic devices.

## Acknowledgements

MS would like to acknowledge support from EPSRC UK grant EP/I017569/1.

# Supporting Information

## Electrical analysis of hysteresis in solution processed silicon nanowire field effect transistors

K. Prabha Rajeev, C. Opoku, V. Stolojan, M. Constantinou and M. Shkunov[*]

Electronic Engineering, Advanced Technology Institute, University of Surrey, Guildford GU2 7XH, UK

* email: m.shkunov@surrey.ac.uk

## Deposition and alignment of silicon nanowires

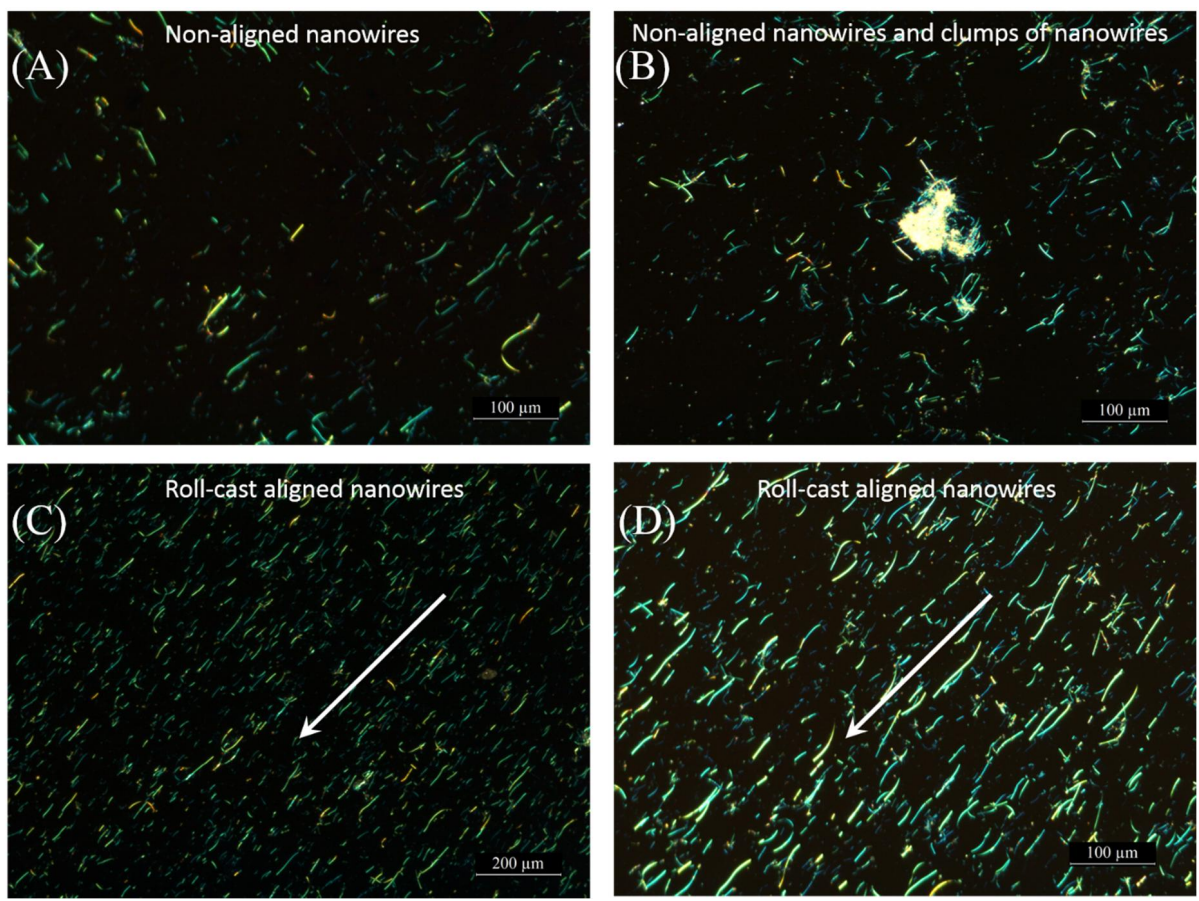

**Figure S1.** Polarised optical microscope images of SFLS grown silicon nanowire on Si/SiO$_2$ substrates: (A) non-aligned nanowires deposited by drop-casting, (B) non-aligned nanowires, with clumps of nanowires on the substrate after drop-casting deposition, (C) and (D) roll-cast aligned nanowires with no clumps of nanowires/particles, arrow represents the direction of alignment.

## Hysteresis and $I_{ON}/I_{OFF}$ calculations

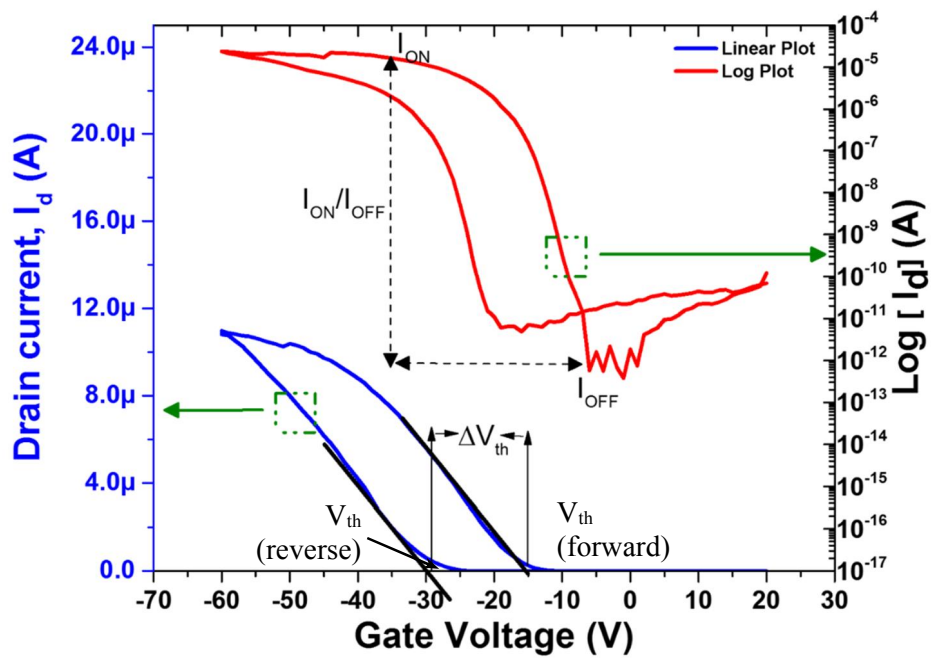

**Figure S2.** Linear regime transfer characteristic of a typical Si-NW FET in a linear and a log-linear scale. Forward Vg scan is from +20V to -60V, immediately followed by a reverse Vg scan from -60V to +20V. Hysteresis ΔVth is extracted from the linear plot by taking the difference between the threshold voltages for the forward and for the reverse scans. Threshold voltage is defined as the intersect of a linear fit to the I-V curves and the zero-current axis. $I_{ON}/I_{OFF}$ current ratio is extracted from the log plot, by taking the ratio of maximum current obtained during accumulation and the current when the FET is in the 'off' state.

# Hysteresis measurements for bottom gate SiO$_2$ dielectric FETs and top gate polymer dielectric FETs.

Data for nine devices of each type is shown. The hysteresis values for SiO$_2$ dielectric FETs are in the 15V to 40V range. Devices with Cytop dielectric demonstrate significantly lower hysteresis between 2V and 10 V, with 50% of devices showing ΔVth of less than 3V.

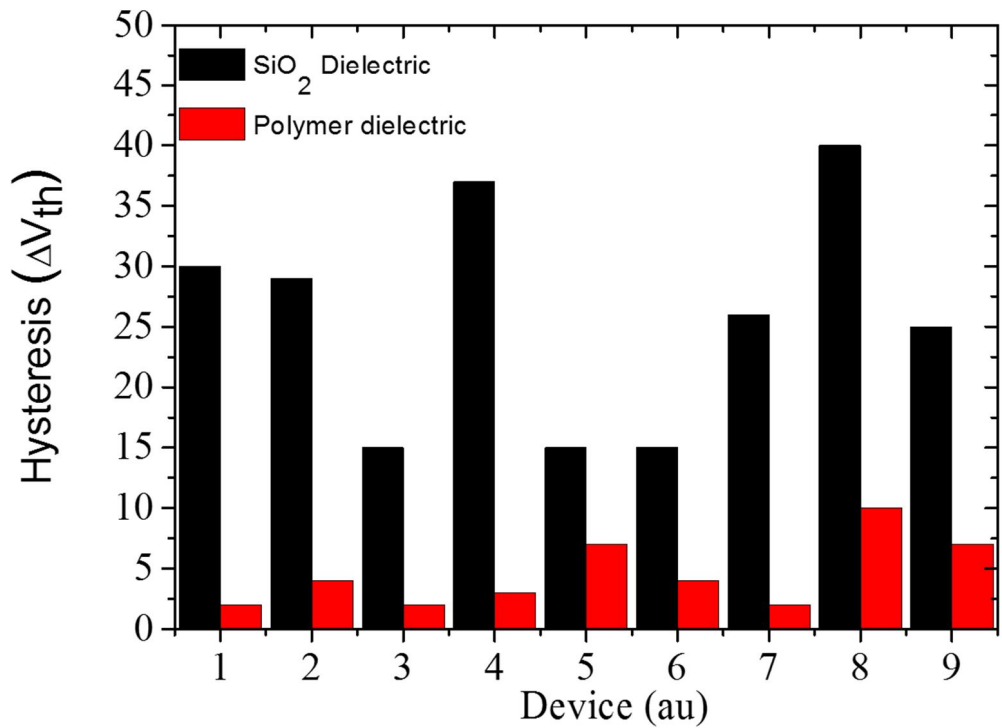

**Figure S3.** The histogram showing the difference in observed hysteresis for various bottom gate Si NW FET with SiO$_2$ dielectric and top gate fluoropolymer Si NW FET. Hysteresis was extracted from transfer I-V measurements conducted at $V_D \sim -6V$.

# Bias stress measurements

**Si nanowire FETs with bottom gate SiO2 dielectric**

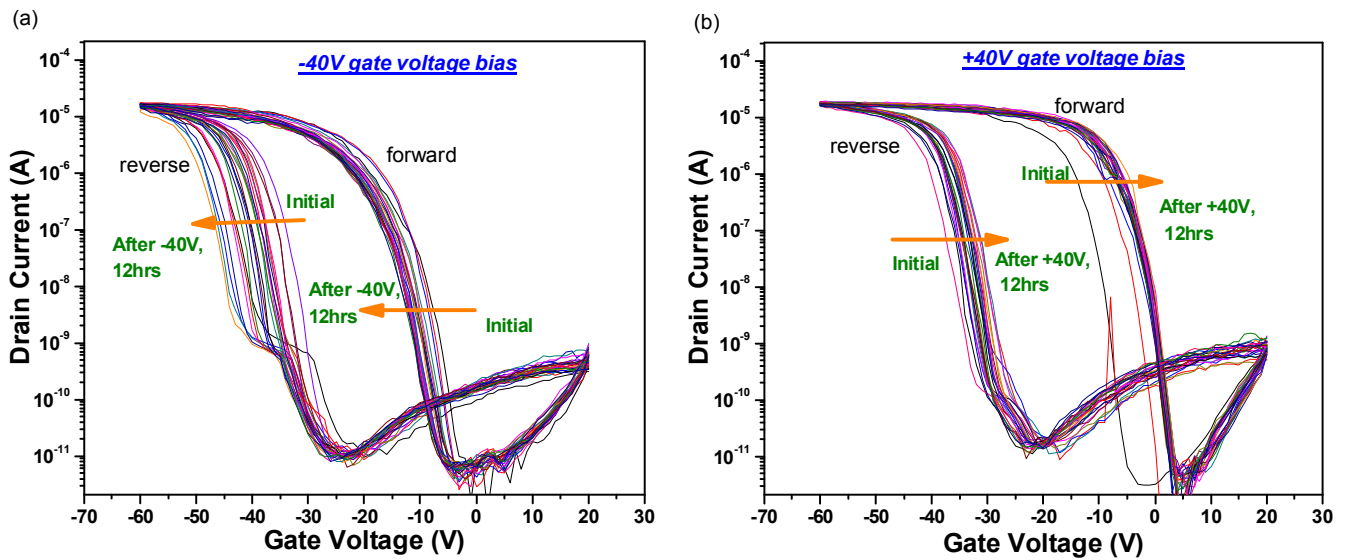

**Figure S4.** Transfer characteristics of bottom gate Si NW FETs with $SiO_2$ dielectric.
(a) transfer characteristics extracted during the gate voltage bias stress measurements with -40 V gate voltage stress, when Vg=-40V applied for 30min & I-Vs measured immediately after, followed by the next cycle of bias stress. Repeated cycles of stress/measurements continued for a total for 12hrs of stress time.
(b) transfer characteristics extracted during the gate voltage bias stress measurements with +40V gate voltage stress, when Vg=+40V applied for 30min & I-Vs measured immediately after, followed by the next cycle of bias stress. Repeated cycles of stress/measurements continued for a total for 12hrs of stress time.
We note that a shift in threshold voltage towards more negative values is observed after negative gate-bias stress, whereas the threshold voltage shifted to positive voltages after positive gate-bias stress.

# Bias stress measurements

**Si nanowire FETs in top gate configuration with Cytop dielectric**

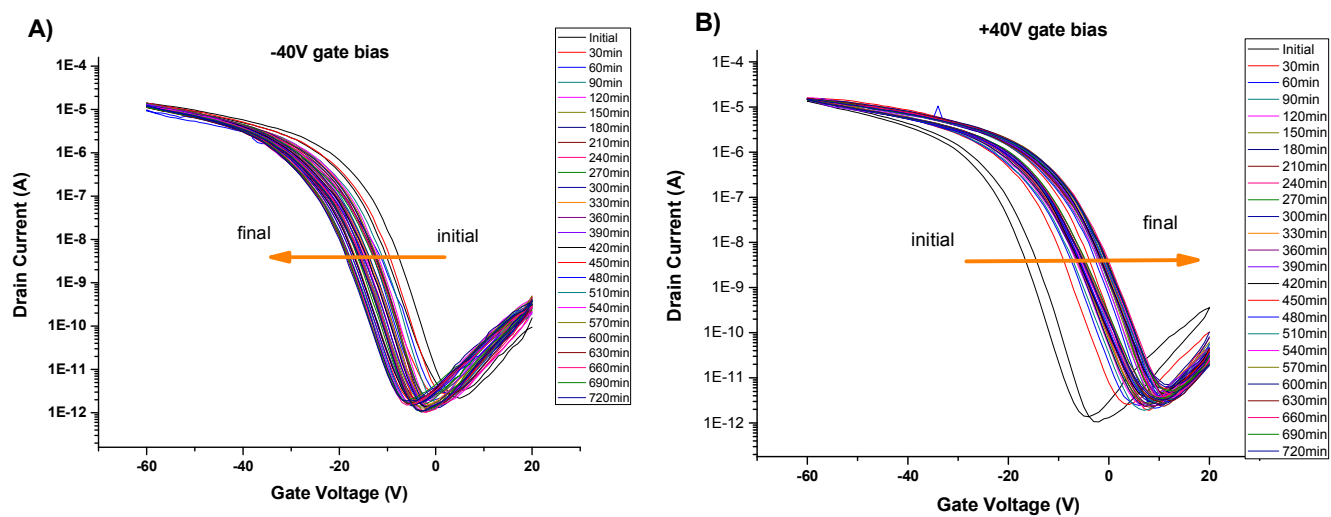

**Figure S5.** Gate voltage bias stress measurements of top gate device with fluoropolymer Cytop dielectric. I-Vs measured at $V_D$=-10V.
(a) Gate bias voltage stress at -40V with a hold time of 30min followed by measurements, cycles repeated for 12hrs. A shift in threshold to more negative voltages is observed.
(b) Gate bias voltage stress at +40V with a hold time of 30min followed by measurement, cycles repeated for 12hrs. Data shows a shift in threshold to the positive gate voltages after the first stress time, but the positive threshold voltage shift is smaller for the next measurements.

## Dual gate Si nanowire devices

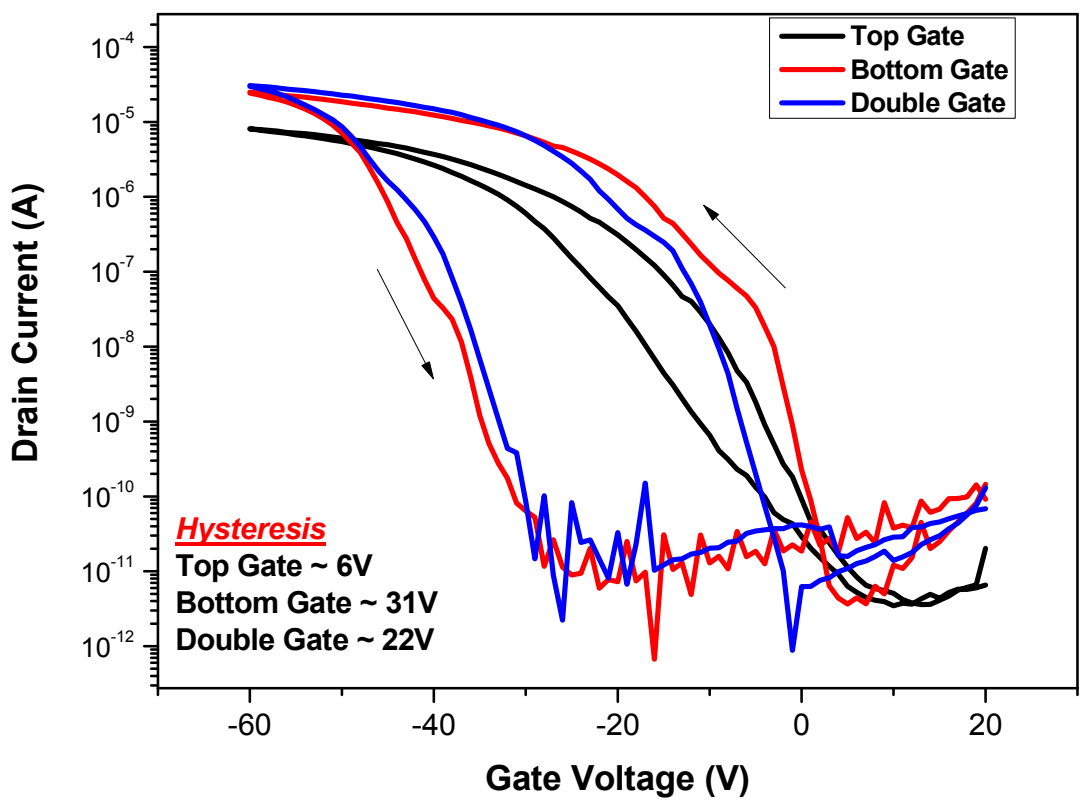

**Figure S6.** Transfer characteristics of the double gated Si nanowire transistor showing individual gate bias characteristics as well as the double gate bias characteristics: red curve – only bottom gate $SiO_2$ is biased, black curve – only top polymer gate is biased, blue curve – both top and bottom gates are biased simultaneously.
Results: ~ 6V hysteresis for Si NW/organic dielectric interface and ~ 31V hysteresis for Si NW/$SiO_2$ interface. When both the gates are biased together, hysteresis reduced to ~ 22V when compared to Si NW/$SiO_2$ interface, and a higher ON current is also observed, compared to top-gate operation.

Dual gate Si NW FET bias stress measurements: each interface (nanowire SiO$_2$) and nanowire/polymer were measured separately.

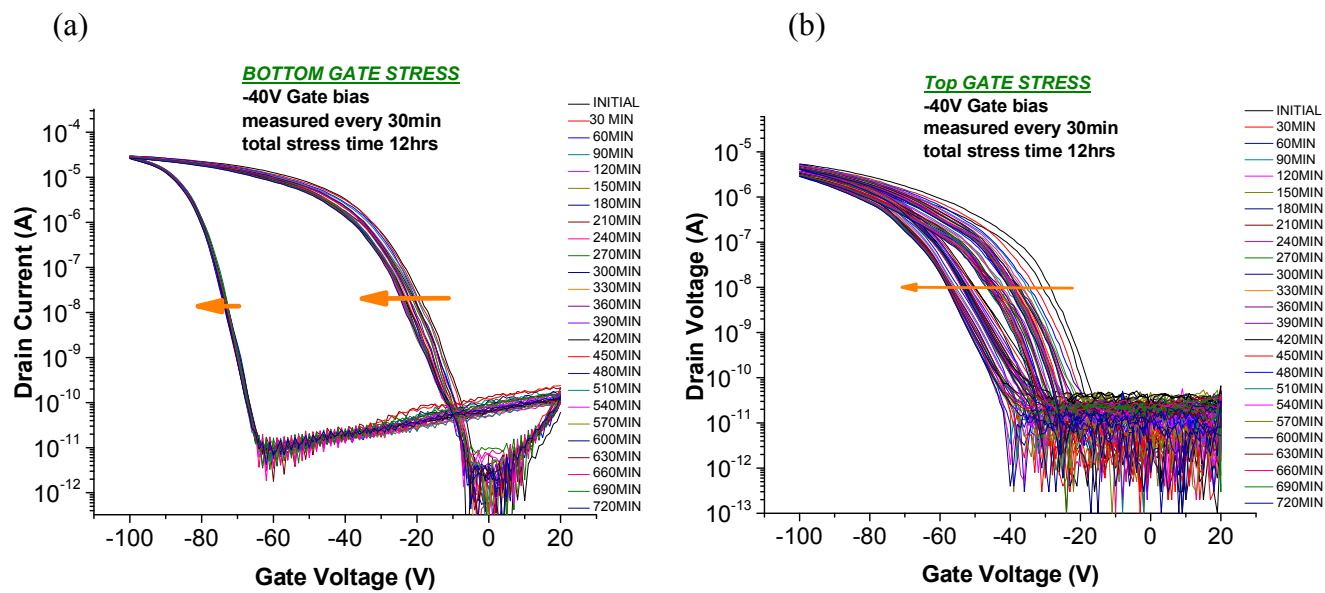

**Figure S7.** Bias stress measurements of dual gate Si NW FET with bottom SiO$_2$ dielectric and top gate fluoropolymer dielectric. (a) Bottom gate device bias voltage stress at -40V with a hold time of 30min followed by measurement, cycles repeated for 12hrs. Data shows a shift in threshold to the negative gate voltages. (b) Top gate device bias voltage stress at -40V with a hold time of 30min followed by measurement, cycles repeated for 12hrs. Results demonstrate a shift in threshold to more negative gate voltages.